\documentclass[]{article}

\usepackage[affil-it]{authblk}
\usepackage{a4wide}
\usepackage{amsmath}
\usepackage{multirow}
\usepackage{booktabs}
\usepackage[ruled,vlined]{algorithm2e}
\usepackage{array}
\usepackage{color}
\usepackage[normalem]{ulem}
\usepackage{hyperref}
\usepackage{amssymb}
\usepackage{graphicx}
\usepackage{amsfonts}
\usepackage[english]{babel}
\usepackage[sort&compress,square,comma,authoryear]{natbib}

\begin{document}
\bibliographystyle{apalike}

\title{Distribution of the H\"{o}gbom CLEAN Algorithm Using Tiled Images with Feedback}

\author[1]{Daniel~Wright}
\author[1]{Karel~Ad\'{a}mek}
\author[1]{Wesley~Armour \thanks{E-mail address: \texttt{wes.armour@oerc.ox.ac.uk}} } 

\affil{Oxford e-Research Centre, Department of Engineering Sciences, University of Oxford, 7 Keble road, OX1 3QG, Oxford, United Kingdom}

\maketitle

\begin{abstract}

Data sizes for next generation radio telescopes, such as the Square Kilometre Array (SKA), are far above that of their predecessors. The CLEAN algorithm was originally developed by \citet{hogbom74}, long before such data sizes were thought possible and is still the most popular tool used for deconvolution in interferometric imaging. In order to facilitate these new large data sizes and reduce computation time a distributed approach to the algorithm has been investigated. The serial nature of the CLEAN algorithm, due to its matching pursuit design, makes this challenging. Splitting the image into a number of tiles which can be individually deconvolved has been investigated, but this creates discontinuities in the deconvolved image and makes it difficult to deconvolve faint sources in the presence of a point spread function associated with bright sources in other tiles. A method of feedback between each of the tiles has been developed to deal with these problems. This new approach has been tested on a simulated dataset containing multiple point sources of known intensity. When compared to a standard H\"{o}gbom deconvolution the tiled feedback version produced a reconstructed image, containing sources up to 2.1 Jy, which agreed to between -0.1 Jy and +0.04 Jy of the standard method across the whole deconvolved image at a speed up to 10.66 times faster.

\end{abstract}

\section{Introduction}

Due to the sparse sampling function of a radio interferometer, deconvolution must be use to recreate the true sky from the measurements taken. By deconvolution the telescopes point spread function (PSF) is removed from the measured image to give the true sky image. The most popular method used to perform this calculation is the CLEAN algorithm which was originally developed by \citet{hogbom74}, this is a matching pursuit algorithm which iteratively attempts to remove fractions of the PSF from the image until a solution is found, so by its nature this is a very sequential algorithm. Due to the massive increase in data sizes for next generation radio telescopes, such as the Square Kilometre Array (SKA), CLEAN has become one of the many bottlenecks in data processing. Here, a method of distributing H\"ogbom CLEAN has been investigated with aim of allowing data to be spread between nodes and speeding up the calculation.

In order to facilitate the investigation a simulated point source dataset was created using the PSF of the VLA in configuration D. The image was created specifically to be challenging, by including a very bright source and because of the unique PSF of the VLA, this makes visualisation of the problem easier. 

\section{Tiled Images}

\begin{figure*}[htb]
     \centering
     \includegraphics[width=.9\textwidth]{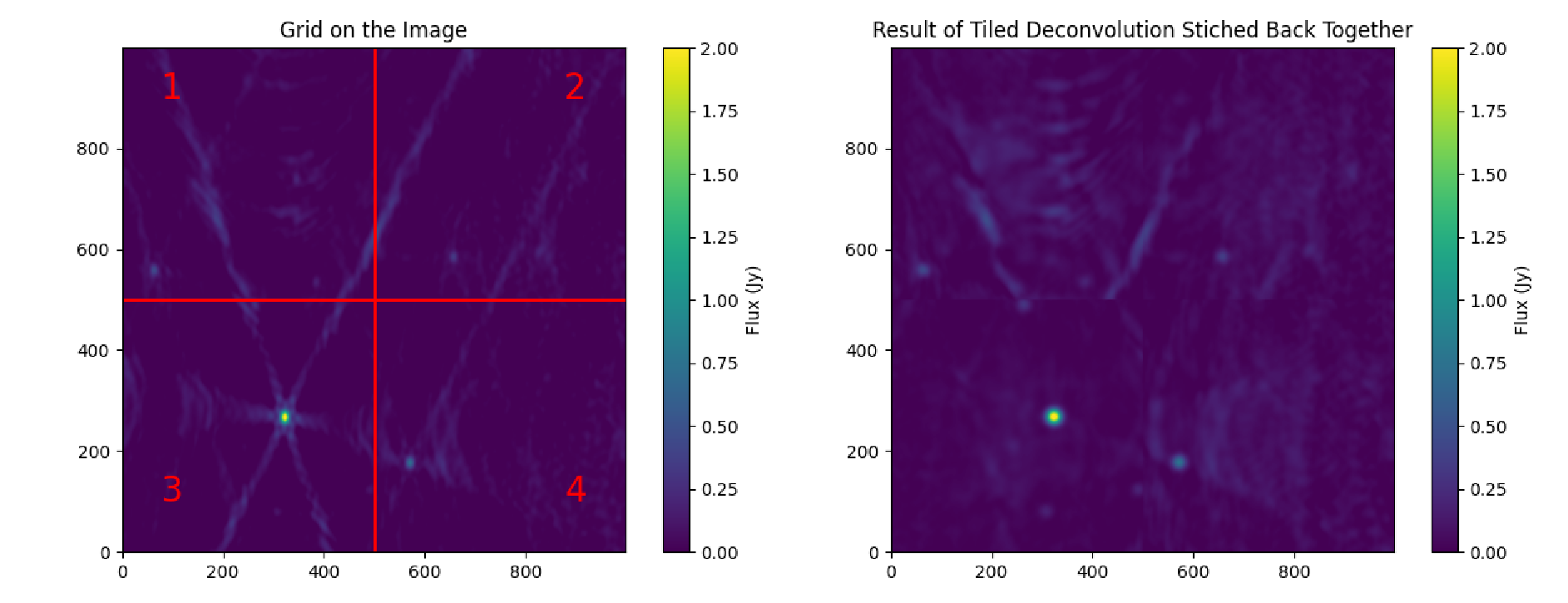}
     \caption{\label{image_grid}(a) Simulated point source Dirty Image with grid overlay (b) Independent tiles deconvolution result}
 \end{figure*}

First, splitting an image in to tiles was investigated. The simulated dirty image is split in to 4 tiles, as shown in Figure~\ref{image_grid}(a), with a separate distributed worker assigned to perform calculations on each tile in parallel.


If each tile is deconvolved independently and the results stitched back together, the result is poor. The lobes of the bright source in tile 3 are not removed from the other tiles, this can be seen in the area of tile 1 in the Figure~\ref{image_grid}(b). Clearly, sources in one tile need to be accounted for in other tiles, hence some form of communication between tiles is needed.

\section{Feedback Between Tiles}

The communication between tiles has been named feedback. The tiles are again deconvolved separately, but then at regular intervals a synchronising feedback step is added to update all the tiles with results from the others

In order to calculate the feedback, some of the common steps of H\"ogbom CLEAN are exploited. In each iteration within a CLEAN minor loop the algorithm finds the pixel within the image with the highest intensity and saves its position and intensity multiplied by gain to a list of CLEAN components:

\begin{displaymath}
I_\mathrm{Components}(l,m) = max(I_\mathrm{Residual}) \times \mathrm{gain}\,.
\end{displaymath}

The PSF is multiplied by the gain and the intensity found and subtracted from the residual image at the location of the highest intensity pixel:

\begin{displaymath}
I_\mathrm{Residual} = I_\mathrm{Residual} - max(I_\mathrm{Residual}) \times \mathrm{gain} \times \mathrm{PSF}\,.
\end{displaymath}

\begin{figure*}[htb]
     \centering
     \includegraphics[width=.9\textwidth]{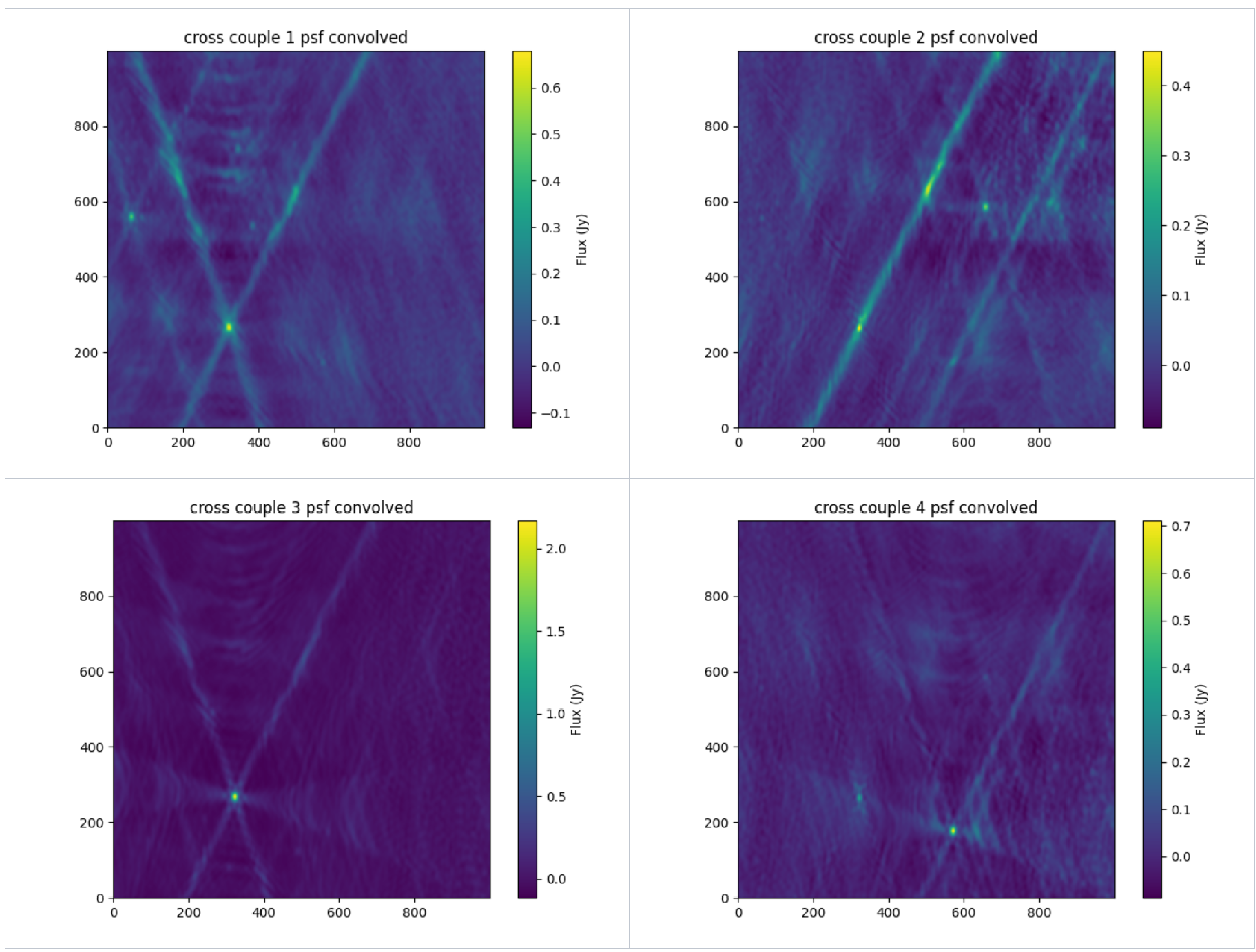}
     \caption{\label{cross_couple}Cross-coupling example for 4 tiles}
 \end{figure*}

Both of these steps act on the position of the highest intensity pixel and both scale by loop gain and intensity. So the CLEAN components for a tile are both a list of where the algorithm thinks a source should be placed and a record of where a PSF was subtracted from. The components can therefore be convolved with the PSF to recreate the total image that has been subtracted from the residual. If each set of components are placed on an image the same size as the full dirty image before being convolved with the full PSF, then not only is the total image subtracted recreated, but also the effects that those subtractions would have had on the other tiles can be seen. Figure~\ref{cross_couple} shows an example of the method, each image shows the effect on the whole image that subtractions in the corresponding tile would have had, if the whole image was being deconvolved. These images have been named cross-coupling.


\begin{figure*}[htb]
     \centering
     \includegraphics[width=.35\textwidth]{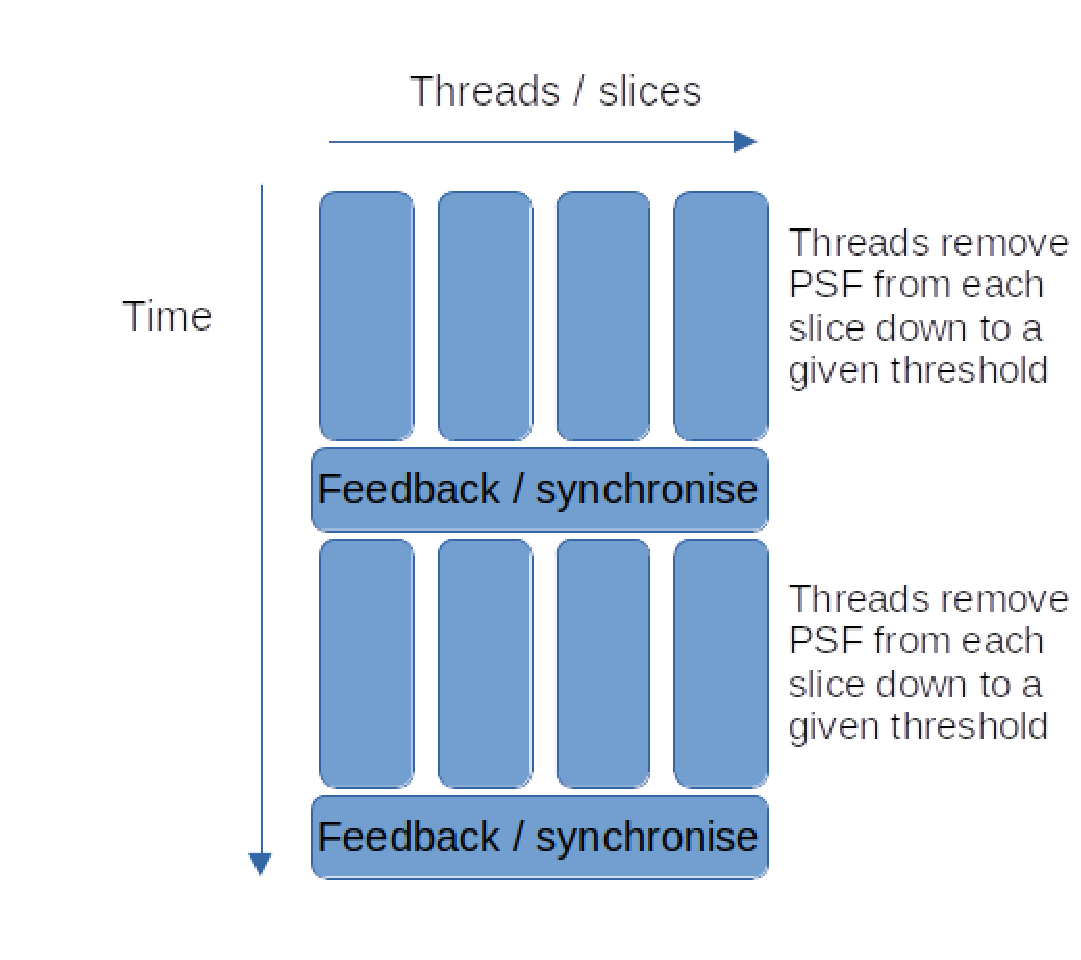}
     \caption{\label{threads_design}Example of multi-step feedback for 4 tiles}
 \end{figure*}

By subtracting each cross-coupling image from the full residual image, the CLEANing in a tile can be propagated to all other tiles. If this feedback technique is used multiple times in the CLEAN process, triggered by meeting a flux threshold, then removal of bright sources can be can be synchronised. This allows their lobes to be removed from other tiles and faint sources to become apparent in those tiles. Figure~\ref{threads_design} shows the overall design of the method.


\section{Results and Conclusions}

\begin{figure*}[htb]
     \centering
     \includegraphics[width=\textwidth]{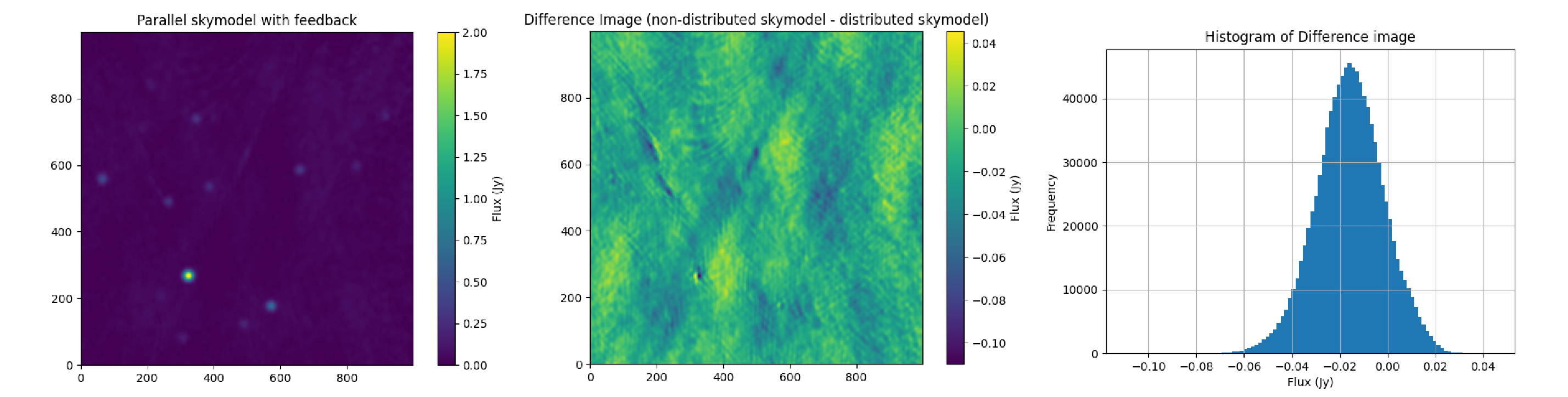}
     \caption{\label{results}(a) Result of a feedback tiled deconvolution (b) difference image to standard method (c) histogram of difference image}
 \end{figure*}

Results are compared to standard Högbom, with the difference between the standard and distributed versions presented. The test dataset contained sources up to 2.1 Jy. The distributed version produced an image which agreed to between -0.1 Jy and +0.04 Jy of the standard method. Results for 4 tiles on a 1024x1024 pixel image are shown in Figure~\ref{results}.


\begin{figure*}[htb]
     \centering
     \includegraphics[width=.65\textwidth]{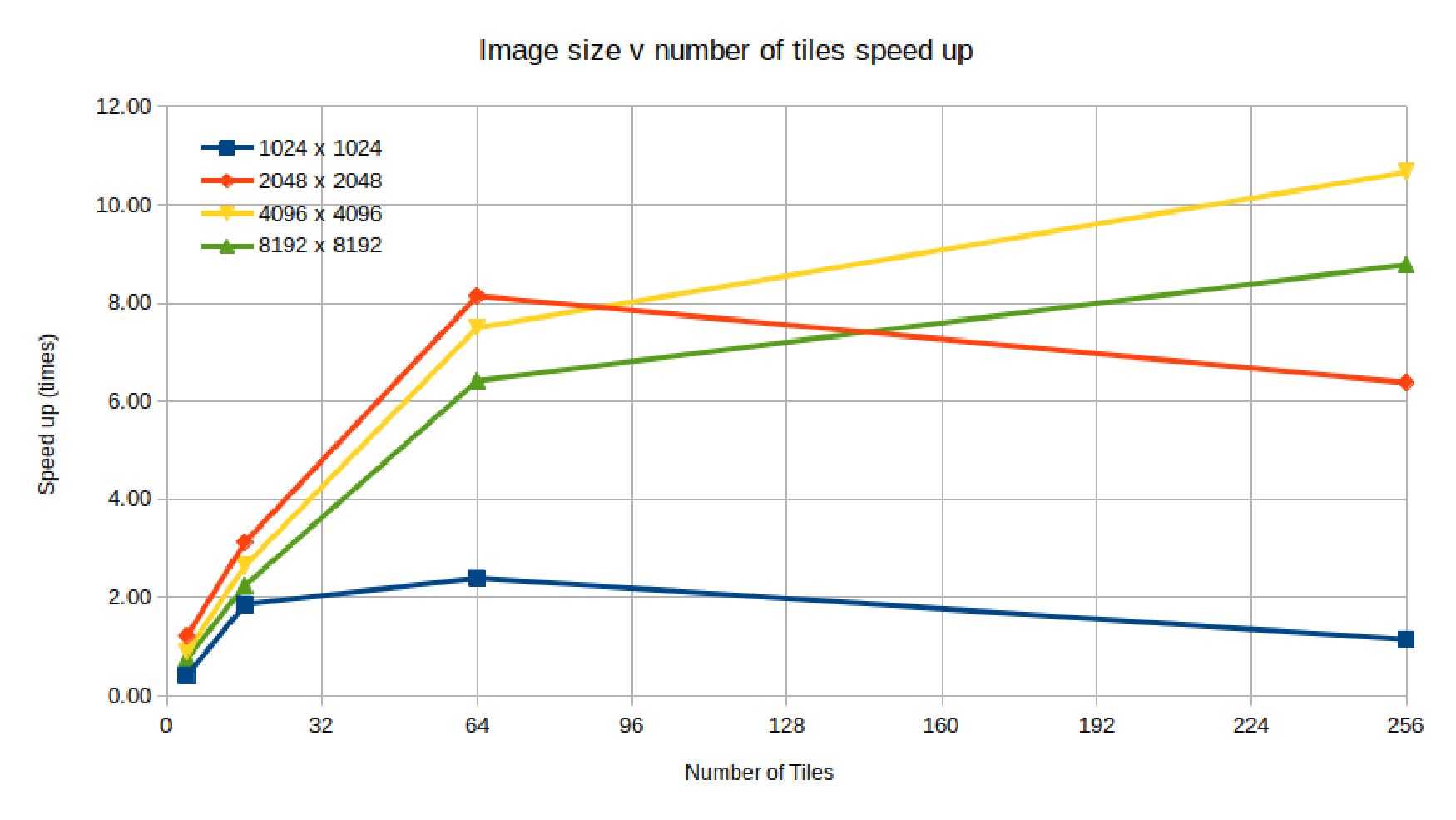}
     \caption{\label{speed_up}Speed up for different image sizes and number of tiles.}
 \end{figure*}

Up to 256 tiles were tested on varying image sizes, using 8 Dask workers on an AMD Ryzen 9 5950x processor. Figure~\ref{speed_up} shows up to a 10.66 times speed up over the standard method was achieved. Tiling beyond 256 was not tested because tiles were becoming too small compared to the size of a source, affecting the quality of the image. The code used can be found at 
\href{https://gitlab.com/dwright550/tiled\_feedback\_hogbom}{https://gitlab.com/dwright550/tiled\_feedback\_hogbom}.


This serves as a proof of concept for this method, allowing the large data volumes of an SKA size telescope to be distributed between nodes, thus reducing the large demands on memory and allowing concurrent processing to speed up calculation time.




\bibliography{p601}  

\section*{Acknowledgements} This work was supported by the Science and Technology Facilities Council [grant number ST/W001969/1].

\end{document}